\RequirePackage{fixltx2e}
\RequirePackage{fix-cm}

\documentclass[%
showkeys,
twocolumn, %
nofootinbib, %
superscriptaddress, %
nofootinbib, %
]{revtex4-1}

\usepackage{cmap}

\usepackage{ucs}
\usepackage[utf8x]{inputenc}
\usepackage[T1,T2A]{fontenc}
\usepackage[german,russian,english]{babel}

\usepackage[sort&compress]{natbib}
\usepackage{amsmath}
\usepackage{amssymb}

\usepackage{mathtools}
\mathtoolsset{
showonlyrefs,
mathic = true
}

\allowdisplaybreaks

\usepackage{hyperref}
\hypersetup{backref,
 colorlinks=false}
\hypersetup{pdfborder=0 0 0}

\usepackage{breakurl}

\usepackage{microtype}
\UseMicrotypeSet[protrusion]{alltext}

\usepackage{fixltx2e}

\makeatletter
\def\ps@pprintTitle{%
     \let\@oddhead\@empty
     \let\@evenhead\@empty
     \let\@oddfoot\@empty
     \let\@evenfoot\@oddfoot}
\makeatother

\DeclareMathOperator{\Rot}{rot}

\newcommand{\setR}{\mathbb{R}}

\newcommand{\crd}[1]{\underline{{#1}}}

\usepackage{tensor}

\begin{document}

\title{Maxwell's Optics Symplectic Hamiltonian}

\author{D. S. Kulyabov}
\email{yamadharma@gmail.com}
\affiliation{Department of Applied Probability and Informatics\\
Peoples' Friendship University of Russia\\
Miklukho-Maklaya str. 6, Moscow, 117198, Russia}
\affiliation{Laboratory of Information Technologies\\
Joint Institute for Nuclear Research\\
Joliot-Curie 6, Dubna, Moscow region, 141980, Russia}

\author{A. V. Korolkova}
\email{avkorolkova@gmail.com} 
\affiliation{Department of Applied Probability and Informatics\\
Peoples' Friendship University of Russia\\
Miklukho-Maklaya str. 6, Moscow, 117198, Russia}

\author{L. A. Sevastyanov}
\email{leonid.sevast@gmail.com}
\affiliation{Department of Applied Probability and Informatics\\
Peoples' Friendship University of Russia\\
Miklukho-Maklaya str. 6, Moscow, 117198, Russia}
\affiliation{Bogoliubov Laboratory of Theoretical Physics\\
Joint Institute for Nuclear Research\\
Joliot-Curie 6, Dubna, Moscow region, 141980, Russia}

\thanks{\emph{Published in:} D. S. Kulyabov, A. V. Korolkova, and
  L. A. Sevastyanov, ``Problem of the Construction of a Hamiltonian
  for Maxwell’s Equations,'' Vestnik natsional’nogo
  issledovatel’skogo yadernogo universiteta ``MIFI'' \textbf{4},
  201--205 (2015), in Russian. 
  \href{http://dx.doi.org/10.1134/S2304487X15030086}{doi:~10.1134/S2304487X15030086}.}

\thanks{\emph{Sources:}
  \url{https://bitbucket.org/yamadharma/articles-2014-em-hamiltonian}}

\begin{abstract}







  The Hamiltonian formalism is extremely elegant and convenient to
  mechanics problems. However, its application to the classical field
  theories is a difficult task.  In fact, you can set one to one
  correspondence between the Lagrangian and Hamiltonian in the case of
  hyperregular Lagrangian.  It is impossible to do the same in 
  gauge-invariant field theories.  In the case of irregular Lagrangian
  the Dirac Hamiltonian formalism with constraints is usually used, and 
  this leads to a number of certain difficulties.  The paper
  proposes a reformulation of the problem to the case of a field
  without sources.  This allows to use a symplectic Hamiltonian
  formalism. The proposed formalism will be used by the authors in the
  future to justify the methods of vector bundles (Hamiltonian
  bundles) in transformation optics.

\end{abstract}


\maketitle

\section{Introduction}
\label{sec:intro}

The Hamiltonian formalism~\cite{luneburg:1964} is well known and widely used 
in geometrical optics. As a Hamiltonian the Hamiltonian
of material particle is used.

In the case of wave optics there is a number of difficulties in the
construction of the Hamiltonian formalism. Since the Lagrangian of the
electromagnetic field is not regular, then the construction of a
symplectic Hamiltonian formalism is not possible. The Dirac formalism with constraints 
is commonly used to gauge theories.

However, if we restrict ourselves to systems without sources, it is
possible to build a standard symplectic Hamiltonian formalism.

This article demonstrates the impossibility of constructing symplectic
Hamiltonian for the general case of the electromagnetic field. In the
case of the sourceless field  the general method for
constructing symplectic Hamiltonian and the example of such constructing are presented.

\section{Notations and conventions}
\label{sec:notation}

  \begin{enumerate}

  \item We will use the notation of abstract
    indices~\cite{penrose-rindler-1987::en}. In this notation tensor
    as a
    complete object is denoted merely by an index (e.g., $x^{i}$). Its
    components are
    designated by underlined indices (e.g., $x^{\crd{i}}$).
    
  \item We will adhere to the following agreements . Greek indices
    ($\alpha$, $\beta$) will refer to the four-dimensional space, in
    component form it looks like: $\crd{\alpha} = \overline{0,3}$.  Latin
    indices from the middle of the alphabet ($i$, $j$, $k$) will refer
    to the three-dimensional space, in the component form it looks like:
    $\crd {i} = \overline{1,3} $.
    
  \item The comma in the index denotes a partial derivative with respect to
    corresponding coordinate ($f_{, i} := \partial_{i}f$); semicolon
    denotes a covariant derivative ($f_{;i} := \nabla_{i} f$).

  \item The CGS symmetrical system~\cite{sivukhin:1979:ufn::en} is used for 
  notation of the equations of electrodynamics.


  \end{enumerate}

\section{Maxwell's equations}

Maxwell's equations will be written both via field variables, and
in the gauge-invariant form (in the formalism of bundles)~\cite{kulyabov:2012:vestnik:2012-1, kulyabov:2013:springer:cadabra,
  kulyabov:2013:conf:maxwell,
  kulyabov:2011:vestnik:curve-maxwell::en,
  kulyabov:2010:conference:slovakiya}.

\subsection{Maxwell's equations via field variables}

We write Maxwell's equations via the field variables in the tensor
formalism in holonomic basis:
\begin{equation}
\label{eq:maxwell:fieldvar}
\left\{
\begin{aligned}
\nabla_i D^i&=4 \pi \rho,\\
e^{ijk}\nabla_j H_k-\frac{1}{c}\partial_tD^i&=\frac{4 \pi }{c}j^i,\\
\nabla^i B_i&=0,\\
e_{ijk}\nabla^j E^k+\frac{1}{c}\partial_t B_i&=0.
\end{aligned}
\right.
\end{equation}

Here $e_{ijk}$ is the Levi--Civita tensor (alternating tensor)\footnote{$e_{ijk}=\sqrt{{}^3g}\varepsilon_{ijk}$,
  $e^{ijk}=\frac{1}{\sqrt{{}^3g}}\varepsilon^{ijk}$,
  $e_{\alpha\beta
    \gamma\delta}=\sqrt{-{}^4g}\varepsilon_{\alpha\beta
    \gamma\delta}$,
  $e^{\alpha\beta
    \gamma\delta}=\frac{1}{\sqrt{-{}^4g}}\varepsilon^{\alpha\beta
    \gamma\delta}$.}.

Field functions $\vec{E}$ and $\vec{B}$ can be represented via field
potentials $\varphi$ and~$\vec{A}$:
\begin{equation}
\label{m:poten}
\vec{B}=\Rot \vec{A},\quad \vec{E}=-\nabla\varphi-\frac{1}{c}\partial_t \vec{A},
\end{equation}
or in the index notation:
\begin{equation}
\label{eq:e_durch_a}
\left\{
\begin{aligned}
B^i & = \left(\Rot \vec{A}\right)^i= e^{ikl}\partial_k A_l,\\
E_i &= - \partial_i \varphi - \partial_0 g_{ij}  A^j.
\end{aligned}
\right.
\end{equation}

\subsection{Maxwell's equations in the formalism of bundles}

We introduce the tensor of the electromagnetic field as a curvature on
the tangent bundle:
\begin{equation}
\label{tensor_F}
F_{\alpha\beta} := \partial_\alpha A_\beta - \partial_\beta A_\alpha,
\end{equation}
where the connection $A^{\alpha}$ means the 4-vector
potential $A^\alpha
= (\varphi, \vec{A})$.

We write the tensor \eqref{tensor_F} by components taking into account relation
\eqref{eq:e_durch_a}:
\begin{equation}
\begin{gathered}
F_{0\crd{i}}=\partial_0 A_{\crd{i}} - \partial_{\crd{i}} A_0 =
- \partial_0 A^{\crd{i}} - \partial _i A^0 
= E_{\crd{i}},\\
F_{\crd{ik}}=\partial_{\crd{i}} A_{\crd{k}}-\partial_{\crd{k}} A_{\crd{i}} = 
-\varepsilon_{\crd{ikl}} B^{\crd{l}}.
\end{gathered}
\end{equation}

Similarly, we introduce the Minkowski's tensor (displacement tensor)
$G^{\alpha\beta} := F^{\alpha \beta} - 4 \pi S^{\alpha \beta}$
($S^{\alpha \beta}$ is the polarization--magnetization tensor)~\cite{jackson:classical_electrodynamics::en}.

Thereby the tensors $F_{\alpha \beta}$ and $G^{\alpha \beta}$ have the following
components
\begin{gather}
F_{\crd{\alpha \beta}}=
\begin{pmatrix} 
0 & {E}_1 & {E}_2 & {E}_3 \\ 
-{E}_1 & 0 & -{B}^3 & {B}^2 \\
-{E}_2 & {B}^3 & 0 & -{B}^1 \\ 
-{E}_3 & -{B}^2 & {B}^1 & 0 
\end{pmatrix},
\label{eq:f_ab}
\\
G^{\crd{\alpha \beta}}=
\begin{pmatrix} 
0 & -{D}^1 & -{D}^2 & -{D}^3 \\ 
{D}^1 & 0 & -{H}_3 & {H}_2 \\
{D}^2 & {H}_3 & 0 & -{H}_1 \\ 
{D}^3 & -{H}_2 & {H}_1 & 0 
\end{pmatrix}.
\label{eq:g^ab}
\end{gather}
  Here $E_{\crd{i}}$, $H_{\crd{i}}$ are components of electric and magnetic
  fields intensity vectors; $D^{\crd{i}}$, $B^{\crd{i}}$ are
  components of vectors of electric and magnetic induction.

Let's write the Maxwell equations with the help of electromagnetic
field tensors $F_{\alpha\beta}$ and
$G_{{\alpha}{\beta}}$~\cite{minkowski:1908,stratton:1948::en}:
\begin{gather}
\nabla_{{\alpha}} F_{{\beta}{\gamma}}+ \nabla_{{\beta}}
F_{{\gamma}{\alpha}}+\nabla_{{\gamma}} F_{{\alpha}{\beta}} = 
F_{[\alpha \beta ; \gamma]} = 0,
\label{eq:m:tensor:2}
\\
\nabla_{{\alpha}} G^{{\alpha}{\beta}}=\frac{4 \pi}{c}j^{{\beta}}.
\label{eq:m:tensor}
\end{gather}

\subsection{The Lagrangian of the electromagnetic field}

Let's construct the Lagrangian ${\mathcal {L}}$ explicitly.

The equation \eqref{eq:m:tensor:2} represents the second Bianchi identity,
that is performed by the construction of~\eqref{tensor_F}.
To construct the Lagrangian we only need to use the
equations~\eqref{eq:m:tensor}.

Then the Lagrangian has the form:
\begin{equation}
{\mathcal{L}} = - \frac{1}{16\pi} F_{\alpha\beta} G^{\alpha\beta} -
\frac{1}{c} j_\alpha A^\alpha.
\end{equation}

 The form of Euler--Lagrange equation:
\begin{equation}
\label{eq:lagr_a}
\nabla_{\beta} \frac{\delta {\mathcal{L}}}{\delta A_{;\beta}^{\alpha}} -
\frac{\delta {\mathcal{L}}}{\delta A^{\alpha}}=0.
\end{equation}

\subsection{The problem of constructing the Hamiltonian of the electromagnetic field}

The Hamiltonian (Hamiltonian density) is constructed by Lagrangian with the help of 
Legendre transformations:
\begin{equation}
\label{eq:ham_a}
\mathcal{H}:= p_{\alpha}\dot{A}^{\alpha}-\mathcal{L},
\end{equation}
where $p_{\alpha}$ is momentum density,
$\mathcal {L}$ is Lagrangian.

Because in Hamiltonian formalism all the equations are constructed with
generalized coordinates and momenta,  in order to write down the Hamiltonian
density~\eqref{eq:ham_a} and the corresponding Hamilton equations we need to
express the generalized momentum $p^{\alpha}$ through
velocities $\dot{A}^{\alpha}$ in the~\eqref{eq:lagr_a}:
\begin{equation}
\left\{
\begin{gathered}
\Dot{A}^{\alpha}=\frac{\delta {\mathcal{H}}}{\delta p_{\alpha}},\\
\Dot{p}_{\alpha}=-\frac{\delta {\mathcal{H}}}{\delta A^{\alpha}}.
\end{gathered}
\right.
\end{equation}

This requires that the determinant of the Hessian matrix is nonzero:
\[
\det {\mathbf{H}} ({\mathcal{L}}) \neq 0,
\]
where the elements of the Hessian matrix are:
\[
\{{\mathbf{H}} ({\mathcal{L}})\}_{\crd{\alpha \beta}} =
\frac{\partial^2 {\mathcal{L}}}{\partial \dot{A}^{\crd{\alpha}} \partial \dot{A}^{\crd{\beta}}}. 
\]
But $F_{00}=0$ and $\{{\mathbf{H}} ({\mathcal{L}})\}_{00}
=\dfrac{\partial^2 {\mathcal{L}}}{\partial (\dot{A}^0
  )^2}=0$. Therefore, $\det {\mathbf{H}} ({\mathcal{L}}) =
0$. That means that the Lagrangian is irregular, and the construction of symplectic
Hamiltonian formalism in such a manner is impossible.

\section{Building a symplectic Hamiltonian}

It turns out that in the absence of sources ($j_{\alpha}=0$)
we can construct a symplectic Hamiltonian formalism making the desired
change of variables. As one of the methods, the method of
doubling variables is considered. This method is applicable when the system contains
only generalized variables, and generalized momenta are absent.

\subsection{Doubling of variables method}

Consider the system of $s$ equations:
\begin{equation}
  \label{eq:doublevar:1}
  \Dot{q}^{\crd{n}} = f^{\,\crd{n}}(q^{n}, q^{n}_{;i}, x^{i}, t), \quad
  \crd{n} = \overline{0,s}.
\end{equation}

We define the space $\setR^{2s}$ with the following coordinates:
\begin{equation}
  \xi^{\crd{n}} := q^{\crd{n}}, \quad \xi^{\crd{n}+\crd{s}}:=
  p_{\crd{n}}, \quad \xi^{a} \in \setR^{2s};
  \quad \crd{n} = \overline{0,s}, \quad
  \crd{a}  = \overline{0,2s}.
\end{equation}

We introduce in this space the Poisson bracket:
\begin{equation}
  \label{eq:poisson}
  \begin{gathered}
    \{ A(\xi^{c},t), B(\xi^{c},t) \} = \Omega^{\crd{ab}}
    \frac{\partial A(\xi^{c},t)}{\partial \xi^{\crd{a}}}
    \frac{\partial B(\xi^{c},t)}{\partial \xi^{\crd{b}}}, 
    \\ 
    \Omega^{\crd{ab}} =
    \begin{pmatrix}
      0 & I \\ -I & 0
    \end{pmatrix}, 
    \quad
    \crd{a}, \crd{b}, \crd{c}  = \overline{0,2s}.
  \end{gathered}
\end{equation}

The Hamiltonian is defined as follows:
\begin{equation}
  \label{eq:doublevar:2}
  \mathcal{H} (q^{n}, p_{n}, x^{i}, t) = p_{\crd{n}} f^{\crd{n}}(q^n, q^n_{;i}, x^{i}, t).
\end{equation}

Then the first group of the Hamilton equations will be the same as
the original system~\eqref{eq:doublevar:1}, and the second group
will be as follows:
\begin{equation}
  \Dot{p}_{\crd{n}} = - \frac{\delta \mathcal{H}}{\delta q^{\crd{n}}} =
  - p_{\crd{m}} \frac{\delta f^{\,\crd{m}}}{\delta q^{\crd{n}}}.
\end{equation}

\subsection{The applicability of the method of doubling the variables for the Maxwell equations}

Let's establish that Maxwell's equations without sources satisfy the
condition of applicability of the doubling of variables method. To do this,
we rewrite the equations~\eqref{eq:maxwell:fieldvar} in the form:
\begin{equation}
\label{eq:maxwell:fieldvar:j=0}
\left\{
\begin{aligned}
\partial_t B_i&= - c e_{ijk}\nabla^j E^k, \\
\partial_t D^i &= c e^{ijk}\nabla_j H_k,\\
\nabla^i B_i &=0,\\
\nabla_i D^i &= 0.\\
\end{aligned}
\right.
\end{equation}

It can be seen that the second pair of equations violates the condition of method applicability. 
However, we can show that in the absence of sources these
equations are linearly dependent on the rest of the Maxwell
equations. Let's write first and second equations in component form:
\begin{equation}
\begin{gathered}
\label{eq:maxwell:fieldvar:j=0:1}
\frac{1}{\sqrt{g}}\left[E_{3,2}-E_{2,3}\right]=-\frac{1}{c}\partial_t
B^1,\\
\frac{1}{\sqrt{g}}\left[E_{1,3}-E_{3,1}\right]=-\frac{1}{c}\partial_t
B^2,\\
\frac{1}{\sqrt{g}}\left[E_{2,1}-E_{1,2}\right]=-\frac{1}{c}\partial_t
B^3.
\end{gathered}
\end{equation}
\begin{equation}
\begin{gathered}
\label{eq:maxwell:fieldvar:j=0:2}
\frac{1}{\sqrt{g}}\left[H_{3,2}-H_{2,3}\right]=\frac{1}{c}\partial_t
D^1,\\
\frac{1}{\sqrt{g}}\left[H_{1,3}-H_{3,1}\right]=\frac{1}{c}\partial_t
D^2,\\
\frac{1}{\sqrt{g}}\left[H_{2,1}-H_{1,2}\right]=\frac{1}{c}\partial_t
D^3.
\end{gathered}
\end{equation}

The following two equations are written in the same way:
\begin{multline}
\label{eq:maxwell:fieldvar:j=0:3}
\frac{1}{\sqrt{g}}\partial_i \left(\sqrt{g} B^i\right) 
= {} \\ {} =
\frac{1}{\sqrt{g}} \left[ \partial_1 \left(\sqrt{g} B^1\right)
  + \partial_2 \left(\sqrt{g} B^2\right) + \partial_3 \left(\sqrt{g}
    B^3\right) \right]
=0,
\end{multline}
\begin{multline}
\label{eq:maxwell:fieldvar:j=0:4}
\frac{1}{\sqrt{g}}\partial_i \left(\sqrt{g} D^i\right) 
= {} \\ {} =
\frac{1}{\sqrt{g}} \left[ \partial_1 \left(\sqrt{g} D^1\right)
  + \partial_2 \left(\sqrt{g} D^2\right) + \partial_3 \left(\sqrt{g} D^3\right) \right]
= 0.
\end{multline}

Differentiating both sides of equations
\eqref{eq:maxwell:fieldvar:j=0:1}
and \eqref{eq:maxwell:fieldvar:j=0:2} we
get (assuming that $g$ is time independent):
\begin{equation}
  \label{eq:maxwell:fieldvar:j=0:1a}
\begin{gathered}
  E_{3, 21}-E_{2,31}= -\frac{1}{c}\partial_t \partial_1 \left(\sqrt{g}
    B^1\right),\\
  E_{1, 32}-E_{3,12}= -\frac{1}{c}\partial_t \partial_2 \left(\sqrt{g}
    B^2\right),\\
  E_{2, 13}-E_{1,23}= -\frac{1}{c}\partial_t \partial_3 \left(\sqrt{g}
    B^3\right).
\end{gathered}
\end{equation}
\begin{equation}
    \label{eq:maxwell:fieldvar:j=0:2a}
\begin{gathered}
  H_{3, 21}-H_{2,31}= \frac{1}{c}\partial_t \partial_1 \left(\sqrt{g}
    D^1\right),\\
  H_{1, 32}-H_{3,12}= \frac{1}{c}\partial_t \partial_2 \left(\sqrt{g}
    D^2\right),\\
  H_{2, 13}-H_{1,23}= \frac{1}{c}\partial_t \partial_3 \left(\sqrt{g}
    D^3\right).
\end{gathered}
\end{equation}

Adding the equations~\eqref{eq:maxwell:fieldvar:j=0:1a}
and~\eqref{eq:maxwell:fieldvar:j=0:2a} term by term
we obtain~\eqref{eq:maxwell:fieldvar:j=0:3}
and~\eqref{eq:maxwell:fieldvar:j=0:4}
respectively.

Thus, the system of Maxwell's equations is transformed into the
following reduced system which satisfies the condition of method:
\begin{equation}
\label{eq:maxwell:fieldvar:j=0:reduced}
\left\{
\begin{aligned}
\partial_t B_i&= - c e_{ijk}\nabla^j E^k, \\
\partial_t D^i &= c e^{ijk}\nabla_j H_k.
\end{aligned}
\right.
\end{equation}

\subsection{Example of the Hamiltonian}

We define the constitutive equations:
\begin{equation}
D^i= \varepsilon^i_j(x^{k})E^j,\quad 
B^i=\mu^i_j(x^{k}) H^j.
\end{equation}

We rewrite~\eqref{eq:maxwell:fieldvar:j=0:reduced} as follows
(assuming that the metric is time independent):
\begin{equation}
\left\{
\begin{aligned}
\partial_t E^i &= c (\varepsilon^{-1})^i_l
\frac{1}{\sqrt{{}^3g}}\varepsilon^{ljk} H_{k,j}, \\
\partial_t H^i &=
- c (\mu^{-1})^i_{l} \frac{1}{\sqrt{{}^3g}} \varepsilon^{ljk} E_{k,j}.
\end{aligned}
\right.
\end{equation}

We choose the generalized coordinates in the form:
\begin{equation}
q^n=\left(E^1, E^2, E^3, H^1, H^2, H^3\right)^{T},
\quad \crd{n} = \overline{1,6}.
\end{equation}

The System~\eqref{eq:doublevar:1} takes the following form:
\begin{equation}
\label{eq:maxwell:general}
\left\{
  \begin{aligned}
    \Dot{q}^{\crd{i}} &= f^{\crd{i}}(q^{n}, q^{n}_{;i}, x^{i}, t) = 
    c (\varepsilon^{-1})^{\crd{i}}_{\crd{l}}
        \frac{1}{\sqrt{{}^3g}}\varepsilon^{\crd{ljk}}
    q_{\crd{k+3},\crd{j}}, \\
    \Dot{q}^{\crd{i+3}} &= f^{\crd{i+3}}(q^{n}, q^{n}_{;i}, x^{i}, t)
    = 
    - c (\mu^{-1})^{\crd{i}}_{\crd{l}} \frac{1}{\sqrt{{}^3g}}
    \varepsilon^{\crd{ljk}} q_{\crd{k},\crd{j}}.    
  \end{aligned}
\right.
\end{equation}

We write the Hamiltonian based on the~\eqref{eq:doublevar:2}
and~\eqref{eq:maxwell:general}:
\begin{multline}
  \mathcal{H} (q^{n}, p_{n}, x^{i}, t) = p_{\crd{n}} f^{\crd{n}}(q^n,
  q^n_{;i}, x^{i}, t) =
  {}\\{}=
  p_{\crd{i}} c (\varepsilon^{-1})^{\crd{i}}_{\crd{l}}
        \frac{1}{\sqrt{{}^3g}}\varepsilon^{\crd{ljk}} q_{\crd{k+3},\crd{j}} -
        p_{\crd{i+3}} c (\mu^{-1})^{\crd{i}}_{\crd{l}} \frac{1}{\sqrt{{}^3g}}
    \varepsilon^{\crd{ljk}} q_{\crd{k},\crd{j}}.
\end{multline}

The corresponding system of Hamiltonian equations has the following
form:
\begin{equation}
\left\{
\begin{gathered}
\Dot{q}^n = \frac{\delta {\mathcal{H}}}{\delta p_n}=f^n,  \\
\begin{multlined}
\Dot{p}_n = - \frac{\delta {\mathcal{H}}}{\delta q^n} = 
- p_{m} \frac{\delta f^m}{\delta q^n}
= {} \\ {} =
- p_m \frac{\partial f^m}{\partial q_n} + 
p_m \partial_i\frac{\partial f^m}{\partial q_{,i}^n} =
p_m \partial_i\frac{\partial f^m}{\partial q_{,i}^n}.
\end{multlined}
\end{gathered}
\right.
\end{equation}

To write out the explicit form of the momenta in general is not
possible.

\section{Conclusion}
\label{sec:conclusion}

In geometrical optics the concept of the ideal instrument was formulated over a hundred years ago.  
Based on this concept, in particular, such focusing (two-dimensional or
three-dimensional) objects, as a Luneburg lens, a Maxwell's fisheye
and others were constructed. Prediction and the invention of lasers have
given for researchers the concept of distribution and transformation of
electromagnetic fields satisfying the Maxwell's equations. In the last 10--20
years, the laws of transformation of three-dimensional vector fields  were studied in 
different papers dedicated to transformational optics. However,
the concept of an ideal instrument in these studies remains taken from
geometrical optics. The authors of this paper intend to strictly
formulate the concept of the ideal 
transformation optics device in Maxwell's optics, that implements
fundamental solutions (propagators, Green's function) 
of Maxwell's equations. The results represented are the first step 
towards solving this problem.

In this paper  a formal method of obtaining symplectic
Hamiltonian formalism for Maxwell's equations without sources was constructed. The
authors also hope that represented examples sufficiently clarify the
application of the proposed method.

\begin{acknowledgments}

The work is partially supported by RFBR grants No's 
13-01-00595, 14-01-00628 and 15-07-08795.

\end{acknowledgments}

\bibliographystyle{elsarticle-num}

\bibliography{bib/em-hamiltonian/geom-maxwell,bib/em-hamiltonian/maxwell-curvecoord,bib/em-hamiltonian/maxwell-curvecoord-other,bib/em-hamiltonian/em-hamiltonian}

\begin{thebibliography}{10}
\expandafter\ifx\csname url\endcsname\relax
  \def\url#1{\texttt{#1}}\fi
\expandafter\ifx\csname urlprefix\endcsname\relax\def\urlprefix{URL }\fi
\expandafter\ifx\csname href\endcsname\relax
  \def\href#1#2{#2} \def\path#1{#1}\fi

\bibitem{luneburg:1964}
R.~K. Luneburg, {Mathematical Theory of Optics}, University of California
  Press, Berkeley \& Los Angeles, 1964.

\bibitem{penrose-rindler-1987::en}
R.~Penrose, W.~Rindler, {Spinors and Space-Time: Two-Spinor Calculus and
  Relativistic Fields}, Vol.~1, Cambridge University Press, 1984.

\bibitem{sivukhin:1979:ufn::en}
D.~V. Sivukhin, \href{http://ufn.ru/en/articles/1979/10/g/}{{The international
  system of physical units}}, Soviet Physics Uspekhi 22~(10) (1979) 834--836.
\newblock \href {http://dx.doi.org/10.1070/PU1979v022n10ABEH005711}
  {\path{doi:10.1070/PU1979v022n10ABEH005711}}.
\newline\urlprefix\url{http://ufn.ru/en/articles/1979/10/g/}

\bibitem{kulyabov:2012:vestnik:2012-1}
D.~S. Kulyabov, A.~V. Korolkova, V.~I. Korolkov, {Maxwell’s Equations in
  Arbitrary Coordinate System}, Bulletin of Peoples’ Friendship University of
  Russia. Series ``Mathematics. Information Sciences. Physics''~(1) (2012)
  96--106.
\newblock \href {http://arxiv.org/abs/1211.6590} {\path{arXiv:1211.6590}}.

\bibitem{kulyabov:2013:springer:cadabra}
A.~V. Korol’kova, D.~S. Kulyabov, L.~A. Sevast’yanov, {Tensor computations
  in computer algebra systems}, Programming and Computer Software 39~(3) (2013)
  135--142.
\newblock \href {http://dx.doi.org/10.1134/S0361768813030031}
  {\path{doi:10.1134/S0361768813030031}}.

\bibitem{kulyabov:2013:conf:maxwell}
D.~S. Kulyabov, \href{http://mmcp2013.jinr.ru}{{Geometrization of
  Electromagnetic Waves}}, in: Mathematical Modeling and Computational Physics,
  JINR, Dubna, 2013, p. 120.
\newline\urlprefix\url{http://mmcp2013.jinr.ru}

\bibitem{kulyabov:2011:vestnik:curve-maxwell::en}
D.~S. Kulyabov, N.~A. Nemchaninova, Maxwell’s equations in curvilinear
  coordinates, Bulletin of Peoples’ Friendship University of Russia. Series
  Mathematics. Information Sciences. Physics~(2) (2011) 172--179, in Russian.

\bibitem{kulyabov:2010:conference:slovakiya}
L.~A. Sevastianov, D.~S. Kulyabov, {The system of Hamilton equations for normal
  waves of the electromagnetic field in a stratified anisotropic medium}, in:
  The 12th small triangle meeting of theoretical physics, Institute of
  experimental physics. Slovak academy of sciences, Stak\v{c}\'{\i}n, 2010, pp.
  82--86.

\bibitem{jackson:classical_electrodynamics::en}
J.~D. Jackson, {Classical Electrodynamics}, 3rd Edition, Wiley, 1998.

\bibitem{minkowski:1908}
H.~Minkowski, {Die Grundlagen f\"{u}r die electromagnetischen Vorg\"{a}nge in
  bewegten K\"{o}rpern}, Nachrichten von der Gesellschaft der Wissenschaften zu
  G\"{o}ttingen, Mathematisch-Physikalische Klasse~(68) (1908) 53--111.

\bibitem{stratton:1948::en}
J.~A. Stratton, {Electromagnetic Theory}, MGH, 1941.

\end{thebibliography}


\begin{thebibliography}{10}
\def\selectlanguageifdefined#1{
\expandafter\ifx\csname date#1\endcsname\relax
\else\selectlanguage{#1}\fi}
\providecommand*{\href}[2]{{\small #2}}
\providecommand*{\url}[1]{{\small #1}}
\providecommand*{\BibUrl}[1]{\url{#1}}
\providecommand{\BibAnnote}[1]{}
\providecommand*{\BibEmph}[1]{#1}
\ProvideTextCommandDefault{\cyrdash}{\hbox to.8em{--\hss--}}
\providecommand*{\BibDash}{\ifdim\lastskip>0pt\unskip\nobreak\hskip.2em\fi
\cyrdash\hskip.2em\ignorespaces}

\bibitem{luneburg:1964}
\selectlanguageifdefined{english}
\BibEmph{Luneburg~R.~K.} {Mathematical Theory of Optics}. \BibDash
\newblock Berkeley \& Los Angeles~: University of California Press, 1964.
  \BibDash
\newblock P.~448.

\bibitem{penrose-rindler-1987}
\selectlanguageifdefined{russian}
\BibEmph{Пенроуз~Р., Риндлер~В.} {Спиноры и
  пространство-время. Два-спинорное
  исчисление и релятивистские поля}. \BibDash
\newblock М.~: Мир, 1987. \BibDash
\newblock \CYRT.~1. \BibDash
\newblock 528~{\cyr\cyrs.}

\bibitem{sivukhin:1979:ufn}
\selectlanguageifdefined{russian}
\BibEmph{Сивухин~Д.~В.} {О Международной системе
  физических величин}~//
  \href{http://dx.doi.org/10.3367/UFNr.0129.197910h.0335}{\BibEmph{Успехи
  физических наук}}. \BibDash
\newblock 1979. \BibDash
\newblock \CYRT. 129, {\cyr\textnumero}~10. \BibDash
\newblock {\cyr\CYRS.}~335--338. \BibDash
\newblock URL: \BibUrl{http://ufn.ru/ru/articles/1979/10/h/}.

\bibitem{kulyabov:2012:vestnik:2012-1}
\selectlanguageifdefined{english}
\BibEmph{Kulyabov~D.~S., Korolkova~A.~V., Korolkov~V.~I.} {Maxwell’s
  Equations in Arbitrary Coordinate System}~// \BibEmph{Bulletin of Peoples’
  Friendship University of Russia. Series ``Mathematics. Information Sciences.
  Physics''}. \BibDash
\newblock 2012. \BibDash
\newblock no.~1. \BibDash
\newblock P.~96--106. \BibDash
\newblock arXiv~: 1211.6590.

\bibitem{kulyabov:2013:springer:cadabra}
\selectlanguageifdefined{english}
\BibEmph{Korol’kova~A.~V., Kulyabov~D.~S., Sevast’yanov~L.~A.} {Tensor
  computations in computer algebra systems}~//
  \href{http://dx.doi.org/10.1134/S0361768813030031}{\BibEmph{Programming and
  Computer Software}}. \BibDash
\newblock 2013. \BibDash
\newblock Vol.~39, no.~3. \BibDash
\newblock P.~135--142.

\bibitem{kulyabov:2013:conf:maxwell}
\selectlanguageifdefined{english}
\BibEmph{Kulyabov~D.~S.} {Geometrization of Electromagnetic Waves}~//
  Mathematical Modeling and Computational Physics. \BibDash
\newblock Dubna~: JINR, 2013. \BibDash
\newblock P.~120. \BibDash
\newblock URL: \BibUrl{http://mmcp2013.jinr.ru}.

\bibitem{kulyabov:2013:tver:maxwell}
\selectlanguageifdefined{russian}
\BibEmph{Кулябов~Д.~С., Королькова~А.~В.}
  {Уравнения Максвелла в произвольной
  системе координат}~// \BibEmph{Вестник
  Тверского государственного университета.
  Серия: Прикладная математика}. \BibDash
\newblock 2013. \BibDash
\newblock {\cyr\textnumero} 1 (28). \BibDash
\newblock {\cyr\CYRS.}~29--44.

\bibitem{kulyabov:2011:vestnik:curve-maxwell}
\selectlanguageifdefined{russian}
\BibEmph{Кулябов~Д.~С., Немчанинова~Н.~А.}
  {Уравнения Максвелла в криволинейных
  координатах}~// \BibEmph{Вестник РУДН. Серия
  «Математика. Информатика. Физика»}. \BibDash
\newblock 2011. \BibDash
\newblock {\cyr\textnumero}~2. \BibDash
\newblock {\cyr\CYRS.}~172--179.

\bibitem{kulyabov:2010:conference:slovakiya}
\selectlanguageifdefined{english}
\BibEmph{Sevastianov~L.~A., Kulyabov~D.~S.} {The system of Hamilton equations
  for normal waves of the electromagnetic field in a stratified anisotropic
  medium}~// The 12th small triangle meeting of theoretical physics. \BibDash
\newblock Stak\v{c}\'{\i}n~: Institute of experimental physics. Slovak academy
  of sciences, 2010. \BibDash
\newblock P.~82--86.

\bibitem{jackson:classical_electrodynamics::ru}
\selectlanguageifdefined{russian}
\BibEmph{Джексон~Д.~Д.} {Классическая
  электродинамика}. \BibDash
\newblock 1965. \BibDash
\newblock {\cyr\CYRS.}~702.

\bibitem{minkowski:1908}
\selectlanguageifdefined{german}
\BibEmph{Minkowski~H.} {Die Grundlagen f\"{u}r die electromagnetischen
  Vorg\"{a}nge in bewegten K\"{o}rpern}~// \BibEmph{Nachrichten von der
  Gesellschaft der Wissenschaften zu G\"{o}ttingen, Mathematisch-Physikalische
  Klasse}. \BibDash
\newblock 1908. \BibDash
\newblock {H.}~68. \BibDash
\newblock S.~53--111.

\bibitem{stratton:1948}
\selectlanguageifdefined{russian}
\BibEmph{Стрэттон~Д.~А.} Теория
  электромагнетизма. \BibDash
\newblock М.-Л.: ГИТТЛ, 1948.

\bibitem{terletskiy-rybakov-1990}
\selectlanguageifdefined{russian}
\BibEmph{Терлецкий~Я.~П., Рыбаков~Ю.~П.}
  Электродинамика. \BibDash
\newblock 2-е, перераб. {\cyr\cyri\cyrz\cyrd.} \BibDash
\newblock М.~: Высшая школа, 1990. \BibDash
\newblock 352~{\cyr\cyrs.}

\bibitem{pavlenko:teormech-lec:2002}
\selectlanguageifdefined{russian}
\BibEmph{Павленко~Ю.~Г.} {Лекции по
  теоретической механике}. \BibDash
\newblock ФИЗМАТЛИТ, 2002. \BibDash
\newblock {\cyr\CYRS.}~392. \BibDash
\newblock
  ISBN:~\href{http://isbndb.com/search-all.html?kw=5-9221-0241-9}{5-9221-0241-9}.

\bibitem{pavlenko:teormech-problems:2003}
\selectlanguageifdefined{russian}
\BibEmph{Павленко~Ю.~Г.} {Задачи по
  теоретической механике}. \BibDash
\newblock ФИЗМАТЛИТ, 2003. \BibDash
\newblock {\cyr\CYRS.}~536. \BibDash
\newblock
  ISBN:~\href{http://isbndb.com/search-all.html?kw=5-9221-0302-4}{5-9221-0302-4}.

\end{thebibliography}

\end{document}


\graphicspath{{image/em-hamiltonian/}}

\title{Симплектический гамильтониан максвелловской оптики}

\author{Д. С. Кулябов}
\email{yamadharma@gmail.com}
\affiliation{Кафедра прикладной информатики и теории вероятностей,\\
  Российский университет дружбы народов,\\
  ул. Миклухо-Маклая, д.6, Москва, Россия, 117198}
\affiliation{Лаборатория информационных технологий,\\
Объединённый институт ядерных исследований,\\
ул. Жолио-Кюри 6, Дубна, Московская область, Россия, 141980}

\author{А. В. Королькова}
\email{avkorolkova@gmail.com} 
\affiliation{Кафедра прикладной информатики и теории вероятностей,\\
  Российский университет дружбы народов,\\
  ул. Миклухо-Маклая, д.6, Москва, Россия, 117198}

\author{Л. А. Севастьянов}
\email{leonid.sevast@gmail.com}
\affiliation{Кафедра прикладной информатики и теории вероятностей,\\
  Российский университет дружбы народов,\\
  ул. Миклухо-Маклая, д.6, Москва, Россия, 117198}
\affiliation{Лаборатория теоретической физики,\\
  Объединённый институт ядерных исследований,\\
  ул. Жолио-Кюри 6, Дубна, Московская область, Россия, 141980}

\thanks{\emph{Опубликовано в:} Кулябов~Д.~С., Королькова~А.~В.,
  Севастьянов~Л.~А. Проблема построения гамильтониана полевых
  уравнений Максвелла~// Вестник НИЯУ МИФИ.~--- 2015.~--- Т.~4,
  \textnumero~3.~---
  С.~201--205. \href{http://dx.doi.org/10.1134/S2304487X15030086}{doi:~10.1134/S2304487X15030086}.}

\thanks{\emph{Исходные тексты:}
  \url{https://bitbucket.org/yamadharma/articles-2014-em-hamiltonian}}

\begin{abstract}
  
    






  Гамильтонов формализм представляется крайне элегантным и удобным в
  задачах механики. Однако его применение к классическим полевым
  теориям представляется достаточно трудной задачей. Действительно,
  можно установить однозначное соответствие между гамильтонианом и
  лагранжианом в случае гиперрегулярного лагранжиана, что не
  выполняеся в калибровочно-инвариантных теориях поля. В случае
  нерегулярного лагранжиана применяется обычно гамильтонов формализм
  со связями, использование которого связано с определёнными
  трудностями. В статье предлагается переформулировка задачи для
  случая полей без источников, что позволяет использовать
  симплектический гамильтонов формализм. Предполагаемый формализм
  будет использован авторами в дальнейшем для обоснования методов
  векторных (гамильтоновых) расслоений в трансформационной оптике.

\end{abstract}

  \keywords{уравнения Максвелла; криволинейные координаты;
    симплектическое многобразие; гамильтонов формализм; удвоение
    переменных}

\maketitle

\section{Введение}
\label{sec:intro}

В геометрической оптике известен и широко применяется гамильтонов
формализм~\cite{luneburg:1964}. В качестве гамильтониана используется гамильтониан
материальной частицы. 

В случае же волновой оптики возникает ряд трудностей при
построении гамильтонова формализма. Поскольку лагранжиан
электромагнитного поля не является регулярным, то построение
симплектического гамильтонового формализма не представляется
возможным. Для калибровочных теорий обычно применяют дираковский
формализм со связями.

Однако, если ограничиться рассмотрением только систем без источников,
то возможно построение стандартного симплектического гамильтонового
формализма.  

В данной статье демонстрируется невозможность построения
симплектического гамильтониана для общего случая электромагнитного
поля. Для случая поля без источников строится общий метод построения
симплектического гамильтониана и приводится пример реализации такого
построения.

\section{Обозначения и соглашения}
\label{sec:notation}

  \begin{enumerate}

  \item Будем использовать нотацию абстрактных
    индексов~\cite{penrose-rindler-1987}. В данной нотации тензор как
    целостный объект обозначается просто индексом (например, $x^{i}$),
    компоненты обозначаются подчёркнутым индексом (например,
    $x^{\crd{i}}$).

  \item Будем придерживаться следующих соглашений.  Греческие индексы
    ($\alpha$, $\beta$) будут относиться к четырёхмерному
    пространству и в компонентном виде будут иметь следующие значения:
    $\crd{\alpha} = \overline{0,3}$. Латинские индексы из середины
    алфавита ($i$, $j$, $k$) будут относиться к трёхмерному
    пространству и в компонентном виде будут иметь следующие значения:
    $\crd{i} = \overline{1,3}$.

  \item Запятой в индексе обозначается частная производная по
    соответствующей координате ($f_{,i} := \partial_{i} f$);
    точкой с запятой --- ковариантная производная ($f_{;i} := \nabla_{i}
    f$).

  \item Для записи уравнений электродинамики в работе используется
    система СГС симметричная~\cite{sivukhin:1979:ufn}.


  \end{enumerate}

\section{Уравнения Максвелла}

Будем записывать уравнения Максвелла как через полевые переменные, так
и в калибровочно-инвариантном виде (в формализме расслоений)~\cite{kulyabov:2012:vestnik:2012-1,
  kulyabov:2013:springer:cadabra, kulyabov:2013:conf:maxwell,
  kulyabov:2013:tver:maxwell,
  kulyabov:2011:vestnik:curve-maxwell,
  kulyabov:2010:conference:slovakiya}.

\subsection{Уравнения Максвелла через полевые переменные}

Уравнения Максвелла через полевые переменные в тензорном формализме в
голономном базисе:
\begin{equation}
\label{eq:maxwell:fieldvar}
\left\{
\begin{aligned}
\nabla_i D^i&=4 \pi \rho,\\
e^{ijk}\nabla_j H_k-\frac{1}{c}\partial_tD^i&=\frac{4 \pi }{c}j^i,\\
\nabla^i B_i&=0,\\
e_{ijk}\nabla^j E^k+\frac{1}{c}\partial_t B_i&=0.
\end{aligned}
\right.
\end{equation}

Здесь $e_{ijk}$ --- тензор Леви--Чивиты (альтернирующий
тензор)\footnote{$e_{ijk}=\sqrt{{}^3g}\varepsilon_{ijk}$,
  $e^{ijk}=\frac{1}{\sqrt{{}^3g}}\varepsilon^{ijk}$,
  $e_{\alpha\beta
    \gamma\delta}=\sqrt{-{}^4g}\varepsilon_{\alpha\beta
    \gamma\delta}$,
  $e^{\alpha\beta
    \gamma\delta}=\frac{1}{\sqrt{-{}^4g}}\varepsilon^{\alpha\beta
    \gamma\delta}$.}.

Полевые функции $\vec{E}$ и $\vec{B}$ можно представить через потенциалы
поля $\varphi$ и~$\vec{A}$:
\begin{equation}
\label{m:poten}
\vec{B}=\Rot \vec{A},\quad \vec{E}=-\nabla\varphi-\frac{1}{c}\partial_t \vec{A},
\end{equation}
или в индексной нотации:
\begin{equation}
\label{eq:e_durch_a}
\left\{
\begin{aligned}
B^i & = \left(\Rot \vec{A}\right)^i= e^{ikl}\partial_k A_l,\\
E_i &= - \partial_i \varphi - \partial_0 g_{ij}  A^j.
\end{aligned}
\right.
\end{equation}

\subsection{Уравнения Максвелла в формализме расслоений}

Введём тензор электромагнитного поля как кривизну на касательном
расслоении:
\begin{equation}
\label{tensor_F}
F_{\alpha\beta} := \partial_\alpha A_\beta - \partial_\beta A_\alpha,
\end{equation}
где связность $A^{\alpha}$ имеет смысл 4-вектора потенциала  $A^\alpha
= (\varphi, \vec{A})$.

Распишем тензор \eqref{tensor_F} по компонентам с учётом
\eqref{eq:e_durch_a}:
\begin{equation}
\begin{gathered}
F_{0\crd{i}}=\partial_0 A_{\crd{i}} - \partial_{\crd{i}} A_0 =
- \partial_0 A^{\crd{i}} - \partial _i A^0 
= E_{\crd{i}},\\
F_{\crd{ik}}=\partial_{\crd{i}} A_{\crd{k}}-\partial_{\crd{k}} A_{\crd{i}} = 
-\varepsilon_{\crd{ikl}} B^{\crd{l}}.
\end{gathered}
\end{equation}

Аналогично введём тензор тензор Минковского (тензор смещений)
$G^{\alpha\beta} := F^{\alpha \beta} - 4 \pi S^{\alpha \beta}$
($S^{\alpha \beta}$ есть тензор поляризации--намагничения)~\cite{jackson:classical_electrodynamics::ru}.

Таким образом тензоры $F_{\alpha\beta}$ и $G^{\alpha\beta}$ имеют
следующие компоненты
\begin{gather}
F_{\crd{\alpha \beta}}=
\begin{pmatrix} 
0 & {E}_1 & {E}_2 & {E}_3 \\ 
-{E}_1 & 0 & -{B}^3 & {B}^2 \\
-{E}_2 & {B}^3 & 0 & -{B}^1 \\ 
-{E}_3 & -{B}^2 & {B}^1 & 0 
\end{pmatrix},
\label{eq:f_ab}
\\
G^{\crd{\alpha \beta}}=
\begin{pmatrix} 
0 & -{D}^1 & -{D}^2 & -{D}^3 \\ 
{D}^1 & 0 & -{H}_3 & {H}_2 \\
{D}^2 & {H}_3 & 0 & -{H}_1 \\ 
{D}^3 & -{H}_2 & {H}_1 & 0 
\end{pmatrix}.
\label{eq:g^ab}
\end{gather}
  Здесь $E_{\crd{i}}$, $H_{\crd{i}}$~--- компоненты векторов напряжённости
  электрического и магнитного полей соответственно; $D^{\crd{i}}$,
  $B^{\crd{i}}$~--- компоненты векторов электрической и магнитной
  индукции соответственно.

Запишем уравнение Максвелла через тензоры
электромагнитного поля 
$F_{\alpha\beta}$ и
$G_{{\alpha}{\beta}}$~\cite{minkowski:1908,stratton:1948, terletskiy-rybakov-1990}:
\begin{gather}
\nabla_{{\alpha}} F_{{\beta}{\gamma}}+ \nabla_{{\beta}}
F_{{\gamma}{\alpha}}+\nabla_{{\gamma}} F_{{\alpha}{\beta}} = 
F_{[\alpha \beta ; \gamma]} = 0,
\label{eq:m:tensor:2}
\\
\nabla_{{\alpha}} G^{{\alpha}{\beta}}=\frac{4 \pi}{c}j^{{\beta}}.
\label{eq:m:tensor}
\end{gather}

\subsection{Лагранжиан электромагнитного поля}

Построим лагранжиан (лагранжеву плотность) ${\mathcal{L}}$ в явном
виде. 

Уравнение \eqref{eq:m:tensor:2} представляет собой дифференциальное
тождество Бьянки, то есть выполняется в силу построения
\eqref{tensor_F}. Для построения лагранжиана достаточно использовать
только группу уравнений~\eqref{eq:m:tensor}.

Тогда лагранжиан будет иметь вид:
\begin{equation}
{\mathcal{L}} = - \frac{1}{16\pi} F_{\alpha\beta} G^{\alpha\beta} -
\frac{1}{c} j_\alpha A^\alpha.
\end{equation}

Уравнения Эйлера--Лагранжа имеют вид:
\begin{equation}
\label{eq:lagr_a}
\nabla_{\beta} \frac{\delta {\mathcal{L}}}{\delta A_{;\beta}^{\alpha}} -
\frac{\delta {\mathcal{L}}}{\delta A^{\alpha}}=0.
\end{equation}

\subsection{Проблема построения гамильтониана электромагнитного поля}

Гамильтониан (гамильтонова плотность) строится через лагранжиан
с помощью преобразований Лежандра:
\begin{equation}
\label{eq:ham_a}
\mathcal{H}:= p_{\alpha}\dot{A}^{\alpha}-\mathcal{L},
\end{equation}
где $p_{\alpha}$ --- плотность импульса,
$\mathcal{L}$ --- лагранжиан.

Так как в гамильтоновом формализме все уравнения строятся через
обобщённые координаты и импульсы, то в \eqref{eq:lagr_a} требуется
выразить обобщённый импульс $p^{\alpha}$ через скорости
$\dot{A}^{\alpha}$, чтобы выписать
гамильтонову плотность \eqref{eq:ham_a} и соответствующие ей уравнения
Гамильтона:
\begin{equation}
\left\{
\begin{gathered}
\Dot{A}^{\alpha}=\frac{\delta {\mathcal{H}}}{\delta p_{\alpha}},\\
\Dot{p}_{\alpha}=-\frac{\delta {\mathcal{H}}}{\delta A^{\alpha}}.
\end{gathered}
\right.
\end{equation}

При этом требуется, чтобы детерминант матрицы Гессе (гессиан) был
отличен от нуля:
\[
\det {\mathbf{H}} ({\mathcal{L}}) \neq 0,
\]
где элементы матрицы Гессе: 
\[
\{{\mathbf{H}} ({\mathcal{L}})\}_{\crd{\alpha \beta}} =
\frac{\partial^2 {\mathcal{L}}}{\partial \dot{A}^{\crd{\alpha}} \partial \dot{A}^{\crd{\beta}}}. 
\]
Но $F_{00}=0$ и $\{{\mathbf{H}} ({\mathcal{L}})\}_{00}
=\dfrac{\partial^2 {\mathcal{L}}}{\partial (\dot{A}^0
  )^2}=0$.  Следовательно, $\det {\mathbf{H}} ({\mathcal{L}}) =
0$. То есть лагранжиан нерегулярный, и построение симплектического
гамильтонового формализма в данном случае невозможно. 

\section{Построение симплектического гамильтониана}

Оказывается, что в случае отсутствия источников ($j_{\alpha} =
0$) можно построить симплектический гамильтонов формализм, произведя
необходимую замену переменных. В качестве одного из методов рассмотрим
метод удвоения переменных~\cite{pavlenko:teormech-lec:2002,
  pavlenko:teormech-problems:2003}. 
Данный метод применим в случае, когда
система содержит лишь обобщённые переменные, а обобщённые импульсы
отсутствуют.

\subsection{Метод удвоения переменных}

Рассмотрим систему $s$ уравнений.
\begin{equation}
  \label{eq:doublevar:1}
  \Dot{q}^{\crd{n}} = f^{\,\crd{n}}(q^{n}, q^{n}_{;i}, x^{i}, t), \quad
  \crd{n} = \overline{0,s}.
\end{equation}

Зададим пространство $\setR^{2s}$ со следующими координатами:
\begin{equation}
  \xi^{\crd{n}} := q^{\crd{n}}, \quad \xi^{\crd{n}+\crd{s}}:=
  p_{\crd{n}}, \quad \xi^{a} \in \setR^{2s};
  \quad \crd{n} = \overline{0,s}, \quad
  \crd{a}  = \overline{0,2s}.
\end{equation}

Введём в этом пространстве скобку Пуассона:
\begin{equation}
  \label{eq:poisson}
  \begin{gathered}
    \{ A(\xi^{c},t), B(\xi^{c},t) \} = \Omega^{\crd{ab}}
    \frac{\partial A(\xi^{c},t)}{\partial \xi^{\crd{a}}}
    \frac{\partial B(\xi^{c},t)}{\partial \xi^{\crd{b}}}, 
    \\ 
    \Omega^{\crd{ab}} =
    \begin{pmatrix}
      0 & I \\ -I & 0
    \end{pmatrix}, 
    \quad
    \crd{a}, \crd{b}, \crd{c}  = \overline{0,2s}.
  \end{gathered}
\end{equation}

А гамильтониан определим следующим образом:
\begin{equation}
  \label{eq:doublevar:2}
  \mathcal{H} (q^{n}, p_{n}, x^{i}, t) = p_{\crd{n}} f^{\crd{n}}(q^n, q^n_{;i}, x^{i}, t).
\end{equation}

Тогда первая группа уравнений Гамильтона будет совпадать с исходной
системой~\eqref{eq:doublevar:1}, а вторая группа будет иметь следующий
вид:
\begin{equation}
  \Dot{p}_{\crd{n}} = - \frac{\delta \mathcal{H}}{\delta q^{\crd{n}}} =
  - p_{\crd{m}} \frac{\delta f^{\,\crd{m}}}{\delta q^{\crd{n}}}.
\end{equation}

\subsection{Соответствие методу}

Установим, что уравнения Максвелла без источников удовлетворяют
условию применимости метода удвоения переменных. Для этого перепишем
уравнения~\eqref{eq:maxwell:fieldvar} в следующем виде:
\begin{equation}
\label{eq:maxwell:fieldvar:j=0}
\left\{
\begin{aligned}
\partial_t B_i&= - c e_{ijk}\nabla^j E^k, \\
\partial_t D^i &= c e^{ijk}\nabla_j H_k,\\
\nabla^i B_i &=0,\\
\nabla_i D^i &= 0.\\
\end{aligned}
\right.
\end{equation}

Видно, что вторая пара уравнений нарушает условие применимости
метода. Однако, можно показать, что в случае отсутствия источников эти
уравнения линейно зависят от остальных уравнений Максвелла. Для этого 
запишем первое и второе уравнения в компонентном виде:
\begin{equation}
\begin{gathered}
\label{eq:maxwell:fieldvar:j=0:1}
\frac{1}{\sqrt{g}}\left[E_{3,2}-E_{2,3}\right]=-\frac{1}{c}\partial_t
B^1,\\
\frac{1}{\sqrt{g}}\left[E_{1,3}-E_{3,1}\right]=-\frac{1}{c}\partial_t
B^2,\\
\frac{1}{\sqrt{g}}\left[E_{2,1}-E_{1,2}\right]=-\frac{1}{c}\partial_t
B^3.
\end{gathered}
\end{equation}
\begin{equation}
\begin{gathered}
\label{eq:maxwell:fieldvar:j=0:2}
\frac{1}{\sqrt{g}}\left[H_{3,2}-H_{2,3}\right]=\frac{1}{c}\partial_t
D^1,\\
\frac{1}{\sqrt{g}}\left[H_{1,3}-H_{3,1}\right]=\frac{1}{c}\partial_t
D^2,\\
\frac{1}{\sqrt{g}}\left[H_{2,1}-H_{1,2}\right]=\frac{1}{c}\partial_t
D^3.
\end{gathered}
\end{equation}

И вторые два уравнения распишем аналогичным образом:
\begin{multline}
\label{eq:maxwell:fieldvar:j=0:3}
\frac{1}{\sqrt{g}}\partial_i \left(\sqrt{g} B^i\right) 
= {} \\ {} =
\frac{1}{\sqrt{g}} \left[ \partial_1 \left(\sqrt{g} B^1\right)
  + \partial_2 \left(\sqrt{g} B^2\right) + \partial_3 \left(\sqrt{g}
    B^3\right) \right]
=0,
\end{multline}
\begin{multline}
\label{eq:maxwell:fieldvar:j=0:4}
\frac{1}{\sqrt{g}}\partial_i \left(\sqrt{g} D^i\right) 
= {} \\ {} =
\frac{1}{\sqrt{g}} \left[ \partial_1 \left(\sqrt{g} D^1\right)
  + \partial_2 \left(\sqrt{g} D^2\right) + \partial_3 \left(\sqrt{g} D^3\right) \right]
= 0.
\end{multline}

Дифференцируя обе части уравнений \eqref{eq:maxwell:fieldvar:j=0:1} и
\eqref{eq:maxwell:fieldvar:j=0:2} получаем (считая, что $g$ не зависит
от времени):
\begin{equation}
  \label{eq:maxwell:fieldvar:j=0:1a}
\begin{gathered}
  E_{3, 21}-E_{2,31}= -\frac{1}{c}\partial_t \partial_1 \left(\sqrt{g}
    B^1\right),\\
  E_{1, 32}-E_{3,12}= -\frac{1}{c}\partial_t \partial_2 \left(\sqrt{g}
    B^2\right),\\
  E_{2, 13}-E_{1,23}= -\frac{1}{c}\partial_t \partial_3 \left(\sqrt{g}
    B^3\right).
\end{gathered}
\end{equation}
\begin{equation}
    \label{eq:maxwell:fieldvar:j=0:2a}
\begin{gathered}
  H_{3, 21}-H_{2,31}= \frac{1}{c}\partial_t \partial_1 \left(\sqrt{g}
    D^1\right),\\
  H_{1, 32}-H_{3,12}= \frac{1}{c}\partial_t \partial_2 \left(\sqrt{g}
    D^2\right),\\
  H_{2, 13}-H_{1,23}= \frac{1}{c}\partial_t \partial_3 \left(\sqrt{g}
    D^3\right).
\end{gathered}
\end{equation}

Складывая почленно~\eqref{eq:maxwell:fieldvar:j=0:1a}
и~\eqref{eq:maxwell:fieldvar:j=0:2a}
получаем~\eqref{eq:maxwell:fieldvar:j=0:3}
и~\eqref{eq:maxwell:fieldvar:j=0:4} соответственно. 

Таким образом, система уравнений Максвелла переходит в следующую
редуцированную систему, удовлетворяющую условию метода:
\begin{equation}
\label{eq:maxwell:fieldvar:j=0:reduced}
\left\{
\begin{aligned}
\partial_t B_i&= - c e_{ijk}\nabla^j E^k, \\
\partial_t D^i &= c e^{ijk}\nabla_j H_k.
\end{aligned}
\right.
\end{equation}

\subsection{Пример реализации гамильтониана}

Зададим материальные уравнения:
\begin{equation}
D^i= \varepsilon^i_j(x^{k})E^j,\quad 
B^i=\mu^i_j(x^{k}) H^j.
\end{equation}

Перепишем~\eqref{eq:maxwell:fieldvar:j=0:reduced} в следующем виде
(считая, что метрика не зависит явным образом от времени):
\begin{equation}
\left\{
\begin{aligned}
\partial_t E^i &= c (\varepsilon^{-1})^i_l
\frac{1}{\sqrt{{}^3g}}\varepsilon^{ljk} H_{k,j}, \\
\partial_t H^i &=
- c (\mu^{-1})^i_{l} \frac{1}{\sqrt{{}^3g}} \varepsilon^{ljk} E_{k,j}.
\end{aligned}
\right.
\end{equation}

Выберем обобщённые координаты в виде:
\begin{equation}
q^n=\left(E^1, E^2, E^3, H^1, H^2, H^3\right)^{T},
\quad \crd{n} = \overline{1,6}.
\end{equation}

Система~\eqref{eq:doublevar:1} приобретает вид:
\begin{equation}
\label{eq:maxwell:general}
\left\{
  \begin{aligned}
    \Dot{q}^{\crd{i}} &= f^{\crd{i}}(q^{n}, q^{n}_{;i}, x^{i}, t) = 
    c (\varepsilon^{-1})^{\crd{i}}_{\crd{l}}
        \frac{1}{\sqrt{{}^3g}}\varepsilon^{\crd{ljk}}
    q_{\crd{k+3},\crd{j}}, \\
    \Dot{q}^{\crd{i+3}} &= f^{\crd{i+3}}(q^{n}, q^{n}_{;i}, x^{i}, t)
    = 
    - c (\mu^{-1})^{\crd{i}}_{\crd{l}} \frac{1}{\sqrt{{}^3g}}
    \varepsilon^{\crd{ljk}} q_{\crd{k},\crd{j}}.    
  \end{aligned}
\right.
\end{equation}

Запишем гамильтониан на основе~\eqref{eq:doublevar:2}
и~\eqref{eq:maxwell:general}:
\begin{multline}
  \mathcal{H} (q^{n}, p_{n}, x^{i}, t) = p_{\crd{n}} f^{\crd{n}}(q^n,
  q^n_{;i}, x^{i}, t) =
  {}\\{}=
  p_{\crd{i}} c (\varepsilon^{-1})^{\crd{i}}_{\crd{l}}
        \frac{1}{\sqrt{{}^3g}}\varepsilon^{\crd{ljk}} q_{\crd{k+3},\crd{j}} -
        p_{\crd{i+3}} c (\mu^{-1})^{\crd{i}}_{\crd{l}} \frac{1}{\sqrt{{}^3g}}
    \varepsilon^{\crd{ljk}} q_{\crd{k},\crd{j}}.
\end{multline}

Соответствующая система уравнений Гамильтона имеет вид:
\begin{equation}
\left\{
\begin{gathered}
\Dot{q}^n = \frac{\delta {\mathcal{H}}}{\delta p_n}=f^n,  \\
\begin{multlined}
\Dot{p}_n = - \frac{\delta {\mathcal{H}}}{\delta q^n} = 
- p_{m} \frac{\delta f^m}{\delta q^n}
= {} \\ {} =
- p_m \frac{\partial f^m}{\partial q_n} + 
p_m \partial_i\frac{\partial f^m}{\partial q_{,i}^n} =
p_m \partial_i\frac{\partial f^m}{\partial q_{,i}^n}.
\end{multlined}
\end{gathered}
\right.
\end{equation}

Выписать явный вид импульсов в общем случае не представляется
возможным.

\section{Заключение}
\label{sec:conclusion}

В геометрической оптике более ста лет назад была сформулирована
концепция идеального прибора. На основе этой концепции были построены,
в частности, такие фокусирующие (двумерные или трёхмерные) объекты,
как линза Люнеберга, <<рыбий глаз>> Максвелла и другие. Предсказание и
изобретение лазеров ввели в обиход исследователей распространение и
преобразование электромагнитных полей, удовлетворяющих уравнениям
Максвелла. Законы трансформации таких трёхмерных векторных полей в
последние 10--20 лет изучаются в работах, посвящённых так называемой
трансформационной оптике. Концепция же идеального прибора в этих
работах остаётся заимствованной из геометрической оптики. Авторы
данной работы намерены строго сформулировать концепцию идеального
трансформационного прибора в максвелловской оптике, реализующего
фундаментальные решения (пропагаторы, функции Грина) системы уравнений
Максвелла. Представленные результаты служат первой ступенью на пути
решения данной задачи.

В работе построен формальный метод получения симплектического
гамильтонового формализма для уравнений Максвелла без
источников. Авторы также надеются, что приведённый пример в
достаточной мере поясняет применение предложенного метода.

\def\acknowledgmentsname{Благодарности}

\begin{acknowledgments}

Работа частично поддержана грантами РФФИ
\textnumero\textnumero~13-01-00595, 14-01-00628 и 15-07-08795.

\end{acknowledgments}

\bibliographystyle{gost2008}

\bibliography{bib/em-hamiltonian/geom-maxwell,bib/em-hamiltonian/maxwell-curvecoord,bib/em-hamiltonian/maxwell-curvecoord-other,bib/em-hamiltonian/em-hamiltonian}